\documentclass [12pt,a4paper]{article}
\usepackage[english]{babel}
\usepackage[]{graphicx}
\usepackage{hyperref}

\author{Boris Mark Tylevich}
\title{A Search Relevancy Tuning Method Using Expert Results Content Evaluation}
\begin{document}

%\maketitle
\begin{center}
{\LARGE{A Search Relevancy Tuning Method Using Expert Results
Content Evaluation}} \newline
 \newline
Boris Mark Tylevich \\Chair of System Integration and Management\\
Moscow Institute of Physics and Technology\\Moscow, Russia
\\email:\htmladdnormallink{boris@tylevich.ru}{mailto:boris@tylevich.ru}

\end{center}

\bigskip
\abstract{The article presents an online relevancy tuning method
using explicit user feedback. The author developed and tested a
method of words' weights modification based on search result
evaluation by user. User decides whether the result is useful or
not after inspecting the full result content. The experiment
proved that the constantly accumulated words weights base leads to
better search quality in a specified data domain. The author also
suggested future improvements of the method.}

\section{Introduction}
The volume of information today increases with a great speed. The
classical search methods (by words match, by keywords etc) return
a great amount of texts that often do not meet the user's
subjective needs. As a result, the user faces a new problem -
"search within the search results". This conducts, first, to
ineffective expenditure of time resources and, second, to decrease
of user confidence in information retrieval system. Therefore, it
is necessary to optimize search results taking into consideration
systematized retrospective data. That is, we the need to design an
expert taught system capable to return search results ranked
according with expert evaluation. At the same time, the user must
not be obliged to make any efforts.

Today's relevance optimization methods, widely used by search
engines, such as considering number of link hits and number of
hyperlinks to the document in other documents (such as
PageRank\cite{google}), are appropriate only in systems of "wide
usage" (Internet) and don't take into account peculiar properties
of corporate knowledge bases. Among them are high concentration of
data in certain knowledge domain and, frequently, absence of
hypertext references. Moreover, the average level of competence of
corporate users is higher than the one of the "broad masses", by
which the Internet is mostly presented.

In this article, the author presents theoretical and practical
issues of designing and developing of an expert system that
constantly optimizes search results in corporate knowledge base.

\section{Earlier work}
The general drawback of using traditional methods of expert
evaluation is that the taught system is being explicitly trained
by experts or information architects until some moment and then
stops in its evolution.

One way to dispose of this problem is the usage of implicit user
feedback. The most evident indication of user's interest in a
document from search results is a "click". By this word, we mean
any possible way of selecting the hypertext link to the document
and proceeding to the document. Thorsten Joachims\cite{joachims}
formally justifies the use of click-through data for training and
describes an unbiased approach to evaluating retrieval
performance. Jared Jacobs et al\cite{jacobs} have recently
proposed a training method using click-through data that tunes
parameter weights in real time rather than offline. This makes
their work a next step in solving the problem of relevance
optimization. It is necessary to mention, that the search in
described works was conducted over the unspecified data (Internet)
that cannot be strictly referred to any knowledge domain. In these
works the results evaluation depended on limited information
sources: the judgment whether the document was useful was bases on
its name, words in URL and a short description (which is often a
few first sentences of the document).

The author assumes that in case of a narrow knowledge domain,
which often corresponds with corporate databases, the earlier
described methods have a substantial disadvantage. When using
standard search within one or several similar knowledge domains it
is much more difficult for the user to assume that the document is
useful, because the judgment is based only on the document's names
and description\footnote{Documents in databases often do not have
a readable URL since data storing methods may vary.}. Another
feature of the author's method of search results raking is the
constant training of the system, since mostly all of its users may
be considered as teaching experts. This partly eliminates the
problem of a limited training period. The developed system gives
users a relatively unconstrained option to express their personal
yet competent opinion.

\section{Algorithm}
In the system developed by author, the user evaluates the quality
of the result after reviewing the whole document. Thus, the
propriety of the evaluation increases noticeably.
\subsection{Ranking}
The author suggests the following algorithm of search results
relevancy ranking R$(q)$:
\begin{equation}
\label{eq:one} f(q,d)=s \cdot N(q,d)
\end{equation}
\begin{equation}
\label{eq:two} R(q,d)=\sum_{i=1}^n c_{i} \cdot f_{i}(q_{i},d)
\end{equation}
\begin{equation}
\label{eq:three} R(d)=\sum_{j=1}^Q w_{j} \cdot R_{j}(q_{j},d)
\end{equation}
where $q$ is a single word from the query $Q$, $d$ is a particular
document, $f$ is a function of query word-document pair,
calculated (\ref{eq:one}) as a product of $s$ (weight of a section
from a section set $S$: name, content etc) and $N$ (the number of
occurrences of word $q$ in document $d$). In summing by section
(\ref{eq:two}) the functions $f$ are multiplied by flag $c$, which
corresponds to a user preference whether to  search in the current
section or not. The final value of the document relevance against
the query is calculated by summing the single document relevance
values against single words that are multiplied by each word's
weight $w$ (\ref{eq:three}). The weights of sections and the
initial words' weights are the inputs to proposed algorithm.

The function $f$ may be defined in several ways. For example, we
may use the query result relevance value returned by SQL-query
(though one should take into consideration several peculiarities
of database systems\cite{mysql}). The better version of the
function may take into account the "distance" between query words
in the result document. That is, the number of words between each
pair of query words. The document with less distance between words
has higher relevance. For example, if the query is "federal courts
in Russia", the document containing phrase "federal and district
courts in Russia" has higher relevance than the document
containing "federal government offices and jury courts in central
Russia".

Most of the search engines are designed on Boolean Retrieval
Model\cite{salton}. A system that adheres to this model supports
the formulation of search queries by combining individual search
words with the Boolean operators AND, OR, and NOT. The result
array is combined of documents through the application of
corresponding AND, OR or NOT set operator. For example, the result
array for a search query "data AND mining" will consist of an
intersection of documents from two sets, each of which contain
only one word from the query. Though being relatively effective,
this model has shortcomings\cite{lu}. The "if and only if"
restriction leads to loss of rank order of retrieved documents,
since a document either satisfies, or does not satisfy, a user's
search query. Another shortcoming is the inability of the user to
specify the importance of each search term.

The author used a simplified version of function, as this question
does not play fundamental role in the experiment. Different
versions of function f may vary the number of query results, but
does not influence the considered ranking effect.

\subsection{Word weight}
After reviewing the document content, the user is able to click a
button as a way to evaluate the search result. After that, the
weights $w$ of a set $W$ of all the query words $W^Q$ that the
document contains $W^D$ are updated.
\begin{equation}
\label{eq:weight} W=W^Q \otimes W^D
\end{equation}
The value of weights modification depend on user's level of
competence $U$ and query result position $p$ in array of results
sorted by relevancy rank $R$:
\begin{equation}
\label{eq:weight2} w^{'}=w+\alpha \cdot U \cdot \sqrt{p}
\end{equation}
The factor $\alpha$ is a learning rate. The user's level of
competence $U$ factor $\alpha$ is defined by experimenter.
\begin{figure}[!htp]
\begin{center}

\includegraphics[width=0.5\textwidth]{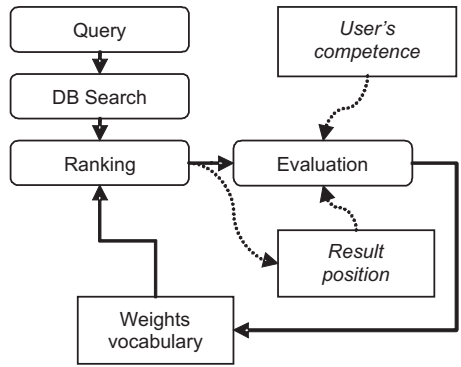}
\caption{The evaluation algorithm diagram} \label{algo}
\end{center}
\end{figure}

Consequently, the set of words with corresponding weights (weights
vocabulary) is updated with every search result evaluation. Figure
\ref{algo} presents the general diagram of the algorithm's work.

It is necessary to note that the method is not applicable to a
search with a single word query, since the ranking algorithm
involves calculating total document's relevancy based on sum of
document's relevancy against each query words. If the query
consists of one word, all of the result documents experience the
same rank alteration. This makes no change in the final results
distribution.

\subsection{Criterion}
After the update of words' weights, it is necessary to recalculate
the rankings of the evaluated documents in search with identical
query. Then a change $\delta$ of current result position is
calculated. The judgment of each search is based on sum $\Delta$
of single position changes:
\begin{equation}
\label{eq:five} \delta=p^{'}-p
\end{equation}
\begin{equation}
\label{eq:six} \Delta=\sum_{i=1}^D \delta_{i}
\end{equation}
where $D$ is quantity of evaluated documents.

\section{Preparation}
The author used his own program to conduct the experiment. The
system developed by author is capable of structurized document
storage and access rights management based on user authentication
information and user groups. The user interacts with the system
via web-interface.

In order to minimize the technical constraints, the system was
installed on Pentium-IV-2.4GHz machine with 2GB of RAM and 160GB
of disk space connected to the Internet with a 2Mb leased line.

The system in the experiment works with texts in Russian.
Therefore, any following information and tables in this article
should be considered as translated from Russian into English. The
morphology of Russian language is more difficult, than English.
For example, the current number of supported word forms for ispell
open-source spellchecker in Russian is 1,24
million\footnote{Lebedev A. Russian dictionary for ispell.
http://scon155.phys.msu.su/~swan/orthography.html}, while in
English it is estimated as less than 1 million. Thus the usage of
Russian does not simplify the examined method.

\subsection{Domain selection}
For the experiment, the author used a special database "The
legislation of Russia" which contained all major legislative
documents of Russian Federation. The content of this base is in
many respects similar to knowledge domains used in corporate
knowledge bases: a substantial quantity of similar documents with
professional terms. Besides, the licensed version of this database
is affordable and each to find. The size of the imported base is
approximately 2,3 millions of words\footnote{The average length of
Russian word with one white space is 6,5 characters. The "clean"
size of imported texts was 15MB.}. The data was imported to the
system manually. The source legislative database was strictly
structured by independent sections of legislation. The author used
the same structure.

\subsection{Technological features}
The system allows conducting a search across several document
fields. For the experiment the author user folder (container)
name, document name and document full text content. The search
engine utilizes a vocabulary of stop-words from MnogoSearch search
engine\cite{mnogo}. Stop-words are widely used and auxiliary
lexical units that are eliminated from the query.

To the each assessor the author assigned a personal level of
competence based on user's subjective estimation of
professionalism in the subject. Each of the assessor was included
in "training experts" user group and given read access to all the
documents in system.

\subsection{Data management methods}
Each search query is saved in database with unique identification
number. Each link in search results contains this ID as well as
the resource (document) unique ID and the number of the result in
result list, i.e. the position of the result in relevance ranking.

The user clicks on document link and opens the document's page,
which contains full document text and embedded dynamic page (using
IFRAME HTML tag). Thus the embedded page has an evaluation submit
form and all the necessary hidden identification information. Upon
evaluation, all this information is submitted to the server. The
server-side program updates all the words that are both in the
query and in the document. The system also saves the current
document position in search results list. After updating the
words' weights, the program conducts an autonomous search with
identical query and saves the difference between the document's
position before and after the evaluation for further analysis.

\section{Experiment}

The initial input values for section weights are set as
$S$(folder, document name, document full text)=$S$(15, 10, 1).
$\alpha$ and $w$ for all words are set equal to 1, as well as all
users' competence level $U$=1.

The relevance judgment logic is binary - whether the document is
relevant to the search query of not. To define relevance for the
assessors, the assessors are told to assume that they are writing
a report on the subject of the query. If they would like to use
any information contained in the document in the report, then the
document should be rated as relevant. The assessors are instructed
to judge a document as relevant regardless of the number of other
documents that contain the same information\cite{TREC2002}. The
assessors are also warned about single word queries. Though such
queries are not prohibited, in these cases the system does not
give the user an option to evaluate the result document. The
topics of the searches are relatively broad and are defined by the
assessors - as much as the imported knowledge base permitted.

\section{Result}
\begin{figure}[!hbp]
\begin{center}

\includegraphics[width=0.5\textwidth]{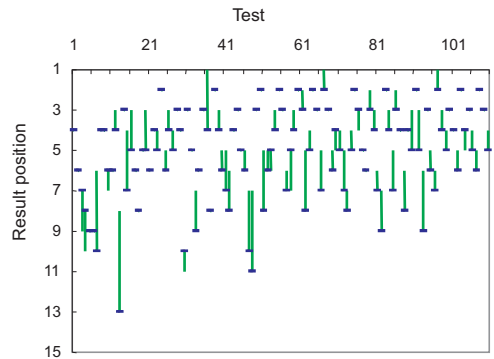}
\caption{Changes in results' positions} \label{results}
\end{center}
\end{figure}

The Figure \ref{results} presents the changes in results'
positions. Blue horizontal bars denote initial result position.
Green vertical bars denote changes in results' positions.

As the experiment indicates, the distribution of results in a
substantial share of tests did not change. Even though in some
tests the position of the result became worse (i.e. increased),
the average effective change in position totalled 0,7. Hence, the
quality of search gradually increased.

\section{Conclusion}
The experiment demonstrates that it is possible to improve the
search results ranking function for a specific knowledge domain
using the method of expert evaluation of results' content and thus
updating words' weights vocabulary.

The main feature of the method is constant improvement of search
quality based on everyday operation. The system is taught every
time the user evaluates the search result if the reviewed document
turns to be useful.

In combination with other technologies of search results
optimization the described methods is able to positively shift the
final results distribution.

\section{Future improvements}
The author supposes that more profound examination of the
algorithm should be conducted, in order to prove right or wrong
the tendency of useful result positions minimalization. The
described experiment has several simplifications.\begin{itemize}
\item It used simplified first-tear relevance function $f$. \item
The search was conducted over one knowledge domain and the initial
values were set empirically. \end{itemize} Therefore, there are
several ways of improving the total effect.

\subsection{Lexical vocabularies}
The experimental system utilized a vocabulary of stop-words in
order to eliminate the search query from useless words. The more
advanced feature is to use vocabularies of synonyms and grammar
forms (stemming). The adoption of these language-specific
vocabularies will upgrade the first-tear search by including more
relevant documents in the result list. The result list then must
be the input to the ranking algorithm described in this article.

\subsection{Division into knowledge domains}
Corporate knowledge bases often contain several knowledge domains,
such as finance and accounting, jurisprudence, technical
documentation, specific data accumulated by employees etc. The
words' significance distribution (i.e. weight vocabulary) for each
of different knowledge domains is unique. Therefore, the next step
in search result optimization is compiling of thematic weight
vocabularies. The compilation and adjustment of vocabularies may
be implemented in many ways: \begin{itemize} \item The user may
independently choose the knowledge domain for his query. The
system uses and updates the corresponding vocabulary. \item The
knowledge administrator can set the vector of competence of the
user a priori. In this case, after every search result evaluation
the system updates all weights vocabularies subject to user's
competence in every knowledge domain.
\end{itemize}

\subsection{Automatic discovery of user competence}
Provided that the system has a substantial volume of accumulated
knowledge - weight vocabularies, it is able to dynamically update
each user's competence vector. The conclusion of quality of each
query may be based on comparison of user's query with each weight
vocabulary. We may suppose that the higher is the total weight of
the query, the higher is user's competence in corresponding
knowledge domain. The user's competence value may become higher or
lower after the comparison of query quality with current user's
competence value. The retrospective analysis of user's competence
vector may be valuable for organization's HR department.

The mentioned suggestions need a profound development and may be
subject for further research.


\begin{thebibliography}{99}
\itemsep=-1pt
\bibitem{salton}
G. Salton and M. McGill. An Introduction to Modern Information
Retrieval, New York, NY: McGraw-Hill, 1983.
\bibitem{lu}
Lu F., Johnsten T., Raghavan V., Traylor D. In Proceedings of the
InForum 99 Conference. Enhancing Internet Search Engines to
Achieve Concept-based Retrieval. Geneva, 1999.
\bibitem{google}
Brin S, Page L. The Anatomy of a Large-scale Hypertextual Search
Engine. Computer Networks and ISDN Systems, 33:107-117, 1998.
\bibitem{joachims}
Joachims, T. Optimizing Search Engines Using Clickthough Data. In
Proceedings of the ACM Conference on Knowledge Discovery and Data
Mining (KDD). Association for Computing Machinery, Edmonton,
Canada, 2002.
\bibitem{jacobs}
Jacobs. J, Rubens. M. An Online Relevancy Tuning Algorithm for
Search Engines.  Stanford Univesity, Stanford, USA, 2003.
\bibitem{mysql}
Golubchik S. MySQL Relevance. In Proceedings of MySQL User
Conference: Fulltext Search, MySQL AB, San Jose, USA, 2003.
\bibitem{mnogo}
http://www.mnogosearch.org/
\bibitem{TREC2002}
Voorhees, E.M. Overview of TREC 2002. In The Eleventh Text
Retrieval Conference (TREC 2002), National Institute of Standards
and Technology, Gaitherburg, USA, 2002.

\end{thebibliography}
\end{document}